\documentclass[twocolumn,showpacs,preprintnumbers,amsmath,amssymb]{revtex4}
\usepackage{amsmath,amssymb,latexsym,graphics,epsfig,subfigure}
\usepackage[T1]{fontenc}

\begin{document}
\renewcommand{\baselinestretch}{1.3}

\title{Charged AdS black hole heat engines}

\author{Shao-Wen Wei \footnote{weishw@lzu.edu.cn},
        Yu-Xiao Liu \footnote{liuyx@lzu.edu.cn}}

\affiliation{Institute of Theoretical Physics, Lanzhou University, Lanzhou 730000, People's Republic of China}

\begin{abstract}
We study the heat engine by a $d$-dimensional charged anti-de Sitter black hole by making a comparison between the small-large black hole phase transition and the liquid-vapour phase transition of water. With the help of the first law and equal-area law, we obtain an exact formula for the efficiency of a black hole engine modeled with a Rankine cycle with or without a back pressure mechanism. When the low temperature is fixed, both the heat and work decreases with the high temperature $T_{1}$. And the efficiency increases with $T_{1}$, while decreases with the charge $q$. For a Rankine cycle with a back pressure mechanism, we find that both the maximum work and efficiency can be approached at the high temperature $T_{1}$. In the reduced parameter space, it also confirms the similar result. Moreover, we observe that the work and efficiency of the black hole heat engine rapidly increase with the number of spacetime dimensions. Thus higher-dimensional charged anti-de Sitter black hole can act as a more efficient power plant producing the mechanical work, and might be a possible source of the power gamma rays and ultrahigh-energy cosmic rays.
\end{abstract}



\maketitle

\section{Introduction}
\label{secIntroduction}

Recently, there is a great interest on studying the anti-de Sitter (AdS) black hole thermodynamics in the extended phase space. The cosmological constant appears to be a new thermodynamic variable \cite{Caldarelli,Cvetic,Lu}, and is interpreted as a pressure  \cite{Kastor,Dolan,Dolan2,Dolan3}
\begin{eqnarray}
 P=\frac{(d-1)(d-2)}{16\pi l^{2}},
\end{eqnarray}
where $d$ is the number of spacetime dimensions and $l$ is the AdS radius. Most importantly, it has been demonstrated that AdS black holes have interesting thermodynamic phenomena and rich phase structure, such as van der Waals (vdW) phase transition, reentrant phase transition, triple point, isolated critical point, and superfluid black hole phase  \cite{Kubiznak,Gunasekaran,Altamirano,Mann,Frassino,Wei0,Kostouki,Wei1,Hennigar} (for a recent review see Ref. \cite{Teo} and references therein). These enhance the fact that a black hole is a thermodynamic system.

Especially, the study of black hole thermodynamics gains a wide generalization by utilizing the phase transition. In our previous work \cite{Wei}, we provided an insight into the black hole microscopic structure by examining the black hole system undergoing a phase transition. The results suggest that the effective microscopic structure of an AdS black hole is similar to an ordinary fluid, and the effective interaction between two molecules, which are introduced to measure the microscopic degrees of freedom of the black hole, can be approximately described by the Lennard-Jones potential function.

Thus an AdS black hole system seems to be a real thermodynamic system, which has the pressure, volume and temperature, and especially is composed of the effective microscopic molecules. Then the physics related to a thermodynamic process starts to get focus. In an isentropic and isobaric process, Dolan \cite{Dolana} first showed that the mechanical work can be yielded from an AdS black hole by decreasing its angular momentum. For example, an extremal electrically neutral or charged rotating black hole can have an efficiency up to 52\% or 75\%. Using the finite-time thermodynamics, the thermodynamic optimization of Penrose process was considered in Ref.~\cite{Bravetti}. The Joule-Thomson expansion for charged AdS black holes was carried out in  Ref. \cite{Aydiner}. Two different kinds of the adiabatic processes were also studied in  Ref. \cite{wenbiao}.

Interestingly, if several thermodynamic processes construct a closed cycle and a black hole operates along it, then a heat engine will be formed. And now the study of the black hole heat engine becomes an important implement of the black hole thermodynamics. Several years ago, Johnson \cite{Johnson} proposed a holographic charged AdS black hole heat engine for the first time. In particular, the thermodynamic cycle was constructed with two isobars and two isochores. The efficiency was worked out in the high temperature limit. Now the study has been extended to other AdS black holes \cite{Belhaj,Sadeghi,Caceres,Setare,Johnson2,Johnson3,Jafarzade,Bhamidipati,
Zhang,Chakraborty,Liang,Hennigar2,LiuMeng,Johnsontb,XuZhao,Yerra,MoLi,Panahiyan}. It is worth noting that an exact efficiency formula was given in Ref.~\cite{Johnson4}.

In general, the working substance in a black hole heat engine is thought to be the black hole fluid, or black hole molecules \cite{Wei}. Thus, as we suggested before, the phase structure and the vdW phase transition of an AdS black hole should be included in. Then, after making a comparison between the small-large black hole phase transition and the liquid-vapour phase transition of water, we investigated the black hole engine along the Rankine cycle along which a steam power plant operates for the first time \cite{WeiLiu}. The result shows that, in the reduced parameter space, the heat, work, and efficiency of the engine are independent of the black hole charge. And the black hole engine working along the Rankine cycle with a back pressure mechanism has a higher efficiency, which could provide us with a novel and efficient mechanism to produce the mechanical work with black hole. Perhaps, the work is used to accelerate the particles near the black hole. Therefore, a black hole engine can act as a possible source of power gamma rays and ultrahigh-energy cosmic rays.

On the other hand, as we know, higher-dimensional charged AdS black holes also exhibit a vdW phase transition \cite{Gunasekaran}. In the reduced parameter space, the critical phenomena were also found to be charge-independent \cite{Wei1}. The aim of this paper is to study the heat engine by $d$-dimensional charged AdS black hole in both the ordinary and reduced parameter spaces.

Our paper is organized as follows. In the next section, we give a brief review of the thermodynamical law and phase transition. In Sec. \ref{Heatengine}, we consider two thermodynamic cycles, the maximal Carnot cycle and the Rankine cycle. By means of the equal-area law, the efficiency for the black hole heat engine operating along these cycles is given. Then we apply our study to the heat engine by a $d$-dimensional charged AdS black hole in Sec. \ref{BHEngine}. The influence of charge and the number of spacetime dimensions on the heat, work, and efficiency is analyzed in detail. In Sec. \ref{Reduced}, we examine the black hole heat engine in the reduced parameter space.
Sec. \ref{Conclusion} is devoted to conclusions and discussions.

\section{Thermodynamical law and phase transition}
\label{law}

Let us first consider the thermodynamical law for an ordinary thermodynamic system containing some substances. During an infinitesimal physical process, the laws can be expressed as
\begin{eqnarray}
 dU&=&TdS-PdV,\\
 dH&=&TdS+VdP.\label{firstlawa}
\end{eqnarray}
where the internal energy $U$ is the amount of energy stored in the system, and the enthalpy $H=U+PV$ measures the energy stored by the heat substances. The variables $T$, $S$, $P$, and $V$ denote the temperature, entropy, pressure, and volume of the system, respectively. All of them are state functions. The heat $Q$ and work $W$ that input into the system from its surroundings can be measured with
\begin{eqnarray}
 \emph{\dj}Q_c&=&TdS,\\
 \emph{\dj}W_c&=&-PdV
\end{eqnarray}
for a closed system, and
\begin{eqnarray}
 \emph{\dj}Q_s&=&TdS,\\
 \emph{\dj}W_s&=&VdP
\end{eqnarray}
for a steady state flow system. The sign $\emph{\dj}$ denotes that the heat $Q$ and the work $W$ are not state functions, which means that $Q$ and $W$ depend on the path. Thus, we have the first law for the closed and steady state flow systems, respectively,
\begin{eqnarray}
 dU=\emph{\dj}Q_c+\emph{\dj}W_c,\\
 dH=\emph{\dj}Q_s+\emph{\dj}W_s.
\end{eqnarray}
Since the internal energy and enthalpy are state functions, one easily has $\Delta U$=0 or $\Delta H$=0 among one thermodynamical cycle, which gives
\begin{eqnarray}
 \Delta Q+\Delta W=0. \label{heatwork}
\end{eqnarray}
Now, let us focus on a special thermodynamic system, the black hole system. Recently, black hole thermodynamics and phase transition were widely studied in the extended phase space, where the cosmological constant was treated as a dynamical pressure. Under such a new interpretation, the cosmological constant can be regarded as a new thermodynamic variable, and its conjugate variable as a thermodynamic volume. For a $d$-dimensional charged AdS black hole, the smarr relation reads
\begin{eqnarray}
 (d-3)M=(d-2)TS-2PV+(d-3)q\Phi.
\end{eqnarray}
Here $M$ is the black hole mass. Differentiating it, we get the black hole first law
\begin{eqnarray}
 dM=TdS+VdP+\Phi dq,\label{firstlaw}
\end{eqnarray}
For fixed charge $q$, the last term vanishes in Eq.~(\ref{firstlaw}). By comparing with Eq.~(\ref{firstlawa}), we see that the black hole mass here should be identified with the enthalpy $M\equiv H$ rather than the internal energy. Then the Gibbs free energy can be obtained by the Legendre transformations: 
\begin{eqnarray}
 G&=&H-TS,\\
 dG&=&-SdT+\Phi dq+VdP.
\end{eqnarray}
For the charged AdS black hole system, it was found that there exists a stable small-large black hole phase transition of vdW type. And the phase transition can be determined by the swallowtail behavior, or equivalently by the equal-area law. The equal-area law was discussed in detail in Ref.~\cite{Spallucci,Wei4} for the charged AdS black holes. For example, in the $T$-$S$ chart, it can be expressed as
\begin{eqnarray}
 \int_{T_{P}}^{T_{P}} SdT=0,
\end{eqnarray}
where $T_{P}$ denotes the temperature of the phase transition. Alternatively, it can also be expressed as
\begin{eqnarray}
 \int_{S_{1}}^{S_{2}} TdS=T_{P}(S_{2}-S_{1}), \label{equalarealaw}
\end{eqnarray}
where $S_{1}$ and $S_{2}$ denote the entropies of the black hole system before and after the phase transition at temperature $T_{P}$.

\section{Heat engine and efficiency}
\label{Heatengine}

A heat engine includes two parts. One is the property of the substances working in the engine, and the other is the thermodynamical cycle along which the engine operates.

As suggested in Ref.~\cite{WeiLiu}, we treat the black hole molecules as the substances. Thus it is a key point to study the phase diagram of the black hole. In fact, the phase structure for the charged AdS black hole was studied in different parameter spaces, and the coexistence curves were numerically obtained in Ref.~\cite{Wei4}. The left thing is the thermodynamical cycle. In this paper, we examine two cycles, the maximum Carnot cycle and the Rankine cycle.

\subsection{Maximal Carnot cycle}

\begin{figure}[!htb]
\center{\subfigure[The Carnot cycle.]{\label{pCarnot1}
\includegraphics[width=7cm]{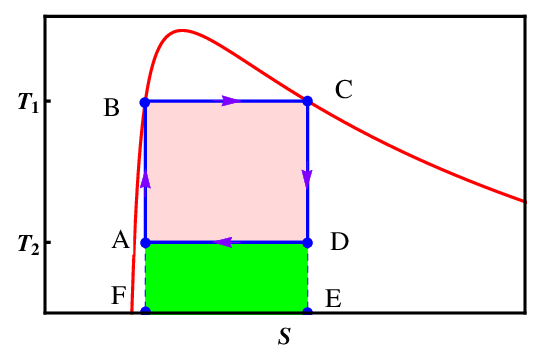}}
\subfigure[The Rankine cycle.]{\label{pRankine1}
\includegraphics[width=7cm]{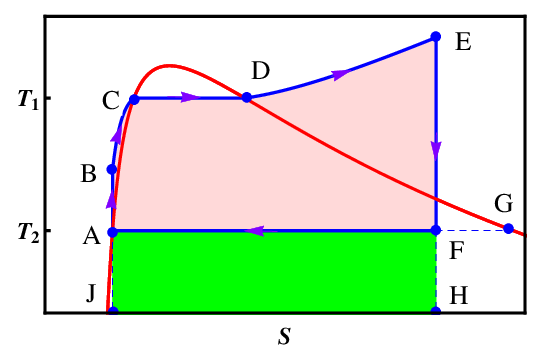}}}
\caption{Sketch pictures of the thermodynamic cycle for black hole heat engines. (a) The Carnot cycle. (b) The Rankine cycle.}\label{skip}
\end{figure}

Carnot cycle is the most simple cycle that contains two temperatures, a heat source and a heat sink. Although the efficiency of a Carnot engine can be directly given, we still like to examine it in a simple way, and it may be useful for studying other thermodynamical cycle. For a charged AdS black hole, one has $S\sim V^{d-2/d-1}$, so the adiabatic line meets the isochoric line, and the Carnot cycle and the Stirling cycle coincide with each other \cite{Johnson}. In order to calculate the efficiency easily, we show the cycle in the $T$-$S$ chart in Fig.~\ref{pCarnot1}. As in Ref.~\cite{WeiLiu}, we here consider the maximal Carnot cycle, which is usually referred to two aspects, during one cycle: i) the substance always works in a coexistence phase; ii) the work output from the engine has a maximum amount. Then, the cycle should be under the coexistence curve, and the states B and C are located on the curve. The Carnot cycle contains four steps, two adiabatic steps (AB and CD) and two isothermal steps (BC and DA). From Fig.~\ref{pCarnot1}, one can clearly see that such cycle is just a rectangle. During one cycle, the black hole engine absorbs the heat $Q_{1}$ form the heat source, and $Q_{2}$ from the heat sink. Meanwhile, the surroundings input the work $W$ to the system. Then, Eq.~(\ref{heatwork}) is reduced to
\begin{eqnarray}
 Q_{1}+Q_{2}+W=0.\label{HW}
\end{eqnarray}
A key quantity to describe the heat engine is the efficiency, which is defined as the ratio of the work output from the machine and the heat input at the higher temperature
\begin{eqnarray}
 \eta=-\frac{W}{Q_{1}}=1+\frac{Q_{2}}{Q_{1}}.
\end{eqnarray}
According to $Q=\int TdS$, the heat absorbed by the system can be measured by the area under the path of a thermodynamical process. Thus, with the help of Fig.~\ref{pCarnot1}, the efficiency can be expressed as
\begin{eqnarray}
 \eta=1-\frac{\text{area(ADEFA)}}{\text{area(BCEFB)}}=1-\frac{T_{2}}{T_{1}}.
\end{eqnarray}
For this cycle, the work done on the system by the surroundings equals to the negative of the area enclosed in the cycle.

\subsection{Rankine cycle}

Among all heat engines operating between the same two temperatures, the reversible Carnot engine has the highest efficiency. However, it is difficult to implement mechanically the kind of engine. One of the practical and extensively used heat engines is the steam power plant, which can be modeled with a Rankine cycle. In each cycle, the working substance, i.e., the water, will undergo a liquid-gas phase transition. Here we would like to consider a black hole heat engine. For clarity, we show the cycle in the $T$-$S$ chart in Fig.~\ref{pRankine1}. The red thick curve represents the coexistence curve. The working substance firstly starts at state A, and approaches state B with increasing temperature and pressure. For simplicity, we regard that this step is adiabatic. Next, the substance follows one isobaric line from state B to state E, and during which, the substance undergoes a phase transition from state C to state D with a constant temperature. Then the substance performs a useful work and reduces its temperature from state E to state F such that the substance is completely in a coexistence phase. Finally, the substance returns to state A by shrinking its volume.

According to the analysis for the Carnot cycle, the efficiency for this Rankine cycle reads
\begin{eqnarray}
 \eta&=&\frac{\text{area(ABCDEFA)}}{\text{area(JBCDEHJ)}}\\
     &=&1-\frac{\text{area(JAFHJ)}}{\text{area(JBCDEHJ)}}.
\end{eqnarray}
The shape of JAFHJ is a rectangle, which can be easily calculated when the states A and F are given. On the other hand, the area of JBCDEHJ includes three parts, the areas under the paths BC, CD, and DE,
\begin{eqnarray}
 \text{area(JBCDEHJ)}&=&\int_{S_{B}}^{S_{C}}TdS+T_{P}(S_{D}-S_{C})\nonumber\\
     & &+\int_{S_{D}}^{S_{E}}TdS.
\end{eqnarray}
As shown above, considering the equal-area law in the $T$-$S$ chart, we have
\begin{eqnarray}
 T_{P}(S_{D}-S_{C})=\int_{S_{C}}^{S_{D}}TdS.
\end{eqnarray}
Therefore, the area of JBCDEHJ reduces to
\begin{eqnarray}
 \text{area(JBCDEHJ)}=\int_{S_{B}(S_{A})}^{S_{E}(S_{F})}TdS.
\end{eqnarray}
Moreover, for fixed charge $q$ and pressure $P$, the first law of the black hole system is $dH_{q,P}=TdS$, or, after integration, 
\begin{eqnarray}
 H_{q,P}=\int TdS.
\end{eqnarray}
Finally, we arrive
\begin{eqnarray}
 \text{area(JBCDEHJ)}=H_{q,P_{B}}(S_{F})-H_{q,P_{B}}(S_{A}).
\end{eqnarray}
Thus, the efficiency is
\begin{eqnarray}
 \eta=1-\frac{T_{A}(S_{F}-S_{A})}{H_{q,P_{B}}(S_{F})-H_{q,P_{B}}(S_{A})}.\label{eff}
\end{eqnarray}
If a back pressure mechanism is introduced for the cycle, the point F will locate at the coexistence curve and coincide with the point G. Then the efficiency reads
\begin{eqnarray}
 \eta=1-\frac{H_{q,P_{A}}(S_{F})-H_{q,P_{A}}(S_{A})}{H_{q,P_{B}}(S_{F})-H_{q,P_{B}}(S_{A})}.
\end{eqnarray}
Since the enthalpy $H$ is pre-given for a black hole system, the calculation will be easily obtained. It is also worth pointing out that the similar formula can be found in Ref.~\cite{Johnson4}. However, the author did not include the phase transition of the black hole \cite{Johnson4}. It is necessary to emphasize that, for a heat engine without the back pressure mechanism, the efficiency can not be easily expressed with the black hole mass, see Eq.~(\ref{eff}).

\section{Black hole heat engine}
\label{BHEngine}

Here, we would like to consider the heat engine by the charged AdS black hole operating along a maximal Carnot cycle or a Rankine cycle.

\subsection{$d$=4-dimensional black holes}

A $d$=4-dimensional charged Reissner-N\"{o}rdstrom AdS black hole is described by the following metric
\begin{eqnarray}
 ds^{2}&=&-f(r)dt^{2}+\frac{dr^{2}}{f(r)}+r^{2}(d\theta^{2}+\sin^{2}\theta d\phi^{2}),\label{metric}\\
 F&=&dA,\quad A=-\frac{q}{r}dt.
\end{eqnarray}
where the metric function is given by
\begin{eqnarray}
 f(r)=1-\frac{2M}{r}+\frac{q^{2}}{r^{2}}+\frac{r^{2}}{l^{2}}.
\end{eqnarray}
The parameters $M$ and $q$ relate to the black hole mass and charge. Such solution  originates from the bulk action
\begin{eqnarray}
 I=-\frac{1}{16\pi}\int d^{4}x\sqrt{-g}\left(R-F^{2}+\frac{6}{l^{2}}\right).
\end{eqnarray}
Through solving $f(r_{h})=0$, we can obtain the radius of the black hole event horizon. Then using the `Euclidean trick', the black hole temperature can be calculated with $T=\frac{\partial_{r}f(r)}{4\pi}|_{r=r_{h}}$:
\begin{eqnarray}
 T=\frac{1}{4\pi r_{h}}\left(1+\frac{3r_{h}^{2}}{l^{2}}-\frac{q^{2}}{r_{h}^{2}}\right).
\end{eqnarray}
The corresponding black hole entropy reads
\begin{eqnarray}
 S=\frac{A}{4}=\pi r_{h}^{2}.
\end{eqnarray}
We can show the temperature and black hole mass (associated with the enthalpy) in terms of the pressure $P$ and entropy $S$,
\begin{eqnarray}
 T&=&\frac{-\pi q^{2}+S+8P S^{2}}{4\sqrt{\pi} S^{3/2}},\\
 H&=&\frac{3\pi q^{2}+3S+8PS^{2}}{6\sqrt{\pi S}}.
\end{eqnarray}
Many works have showed that this black hole system exhibits a small-large black hole phase transition, with the critical point given by \cite{Kubiznak}
\begin{eqnarray}
 T_{c}&=&\frac{\sqrt{6}}{18\pi q},\;\;
 S_{c}=6\pi q^{2},\\
 v_{c}&=&2\sqrt{6}q,\;\;
 P_{c}=\frac{1}{96\pi q^{2}}.\label{criticalpoint}
\end{eqnarray}
In general, there exists no analytical form of the coexistence curve for a thermodynamic system, even for the most simple model, the vdW fluid. However, fortunately, an analytical form of the coexistence curve of small and large black holes was obtained in Ref.~\cite{Lan} by solving the equal-area law in the reduced parameter space. After a simple parameter transformation, one gets
\begin{widetext}
\begin{eqnarray}
 T=\frac{1}{3\sqrt{6}\pi q}
   \sqrt{1+\cos\left(3+\arccos\left(\frac{12\pi q^2+S-6q\sqrt{3\pi^2q^2+\pi S}}{2S}\right)\right)},\label{tscoe}
\end{eqnarray}
\end{widetext}
where we require $S>\pi q^2$.

\begin{figure}
\includegraphics[width=8cm]{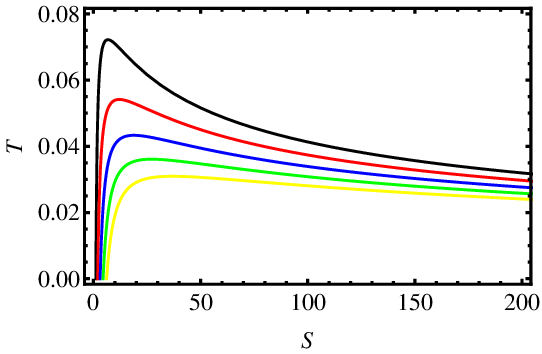}
\caption{Phase diagram in the $T$-$S$ chart. The parameter is set to $q$=0.6, 0.8, 1, 1.2, and 1.4 from top to bottom.} \label{pts}
\end{figure}

We plot the coexistence curve in the $T$-$S$ chart for different values of the charge $q$ in Fig.~\ref{pts}. For each fixed $q$, there are three regions. Below the curve is the coexistence region of small and large black holes. And  above it, they are small black hole region located at the left and large black hole region located at the right. Moreover, it is clear that the peak (related to the critical point) decreases and shifts to large $S$ with the increasing of the black hole charge. Thus increasing $q$ will shrink the coexistence region.

\begin{figure}
\center{\subfigure[Work for the maximal Carnot cycle.]{\label{pCarnota}
\includegraphics[width=7cm]{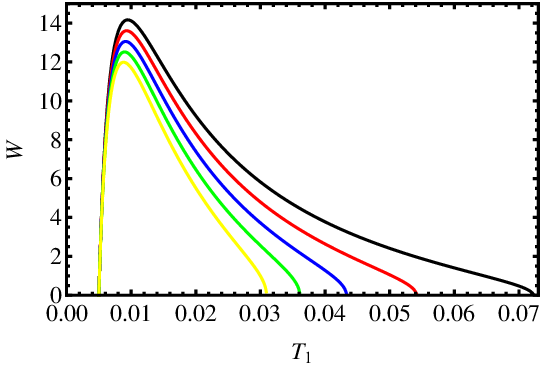}}
\subfigure[Heat for the maximal Carnot cycle.]{\label{pCarnotb}
\includegraphics[width=7cm]{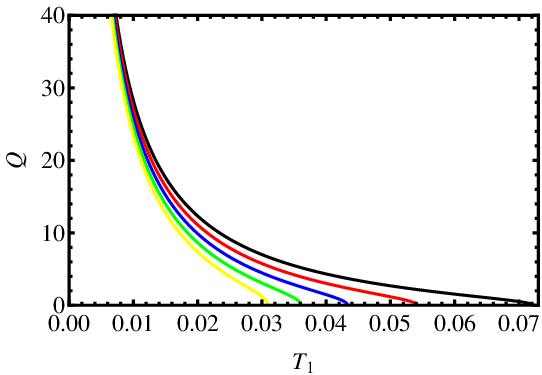}}}
\caption{(a) Work and (b) heat as a function of $T_{1}$ for the maximal Carnot cycle with the low temperature $T_{2}$=0.005, and $q$=0.6, 0.8, 1, 1.2, and 1.4 from top to bottom.}\label{pCarnot}
\end{figure}

Based on this phase structure, we would like to turn to the black hole heat engine. At first, let us focus on the maximal Carnot cycle, which means that the engine always operates in the coexistence region, and points B and C are located on the coexistence curve. In such scheme, we plot the work and heat in Fig.~\ref{pCarnot} with the low temperature $T_{2}$=0.005, and the high temperature $T_{1}$ varies from $T_{2}$ to the critical temperature. For different values of $q$, the work and heat share the similar behavior, respectively. From Fig.~\ref{pCarnota}, it can be seen that the work $W$ first increases from zero at $T_{2}$, then approaches to its maximum, and finally decreases to zero at the critical temperature. The peak of the work decreases with the charge $q$ and slightly shifts to the low temperature. The heat described in Fig.~\ref{pCarnotb} shows a monotonically decreasing behavior. And the larger the charge is, the faster the decrease is. They approach to zero at the critical temperature. Nevertheless, the efficiency of the heat engine is in a compact form, i.e., $\eta=1-\frac{T_{2}}{T_{1}}$.

\begin{figure}
\center{\subfigure[Heat for the Rankine cycle.]{\label{pRankineHeat}
\includegraphics[width=7cm]{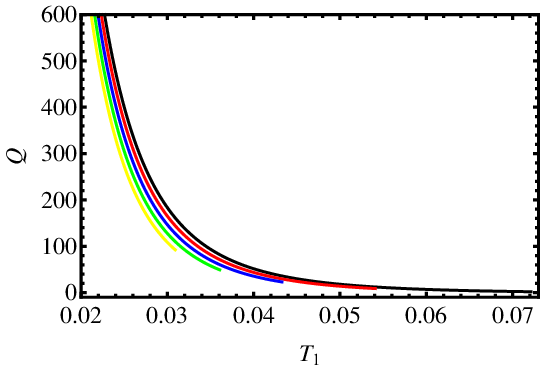}}
\subfigure[Work for the Rankine cycle.]{\label{pRankineWork}
\includegraphics[width=7cm]{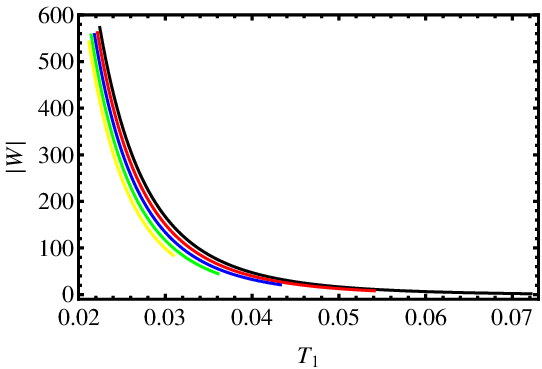}}}
\caption{(a) Heat and (b) work as a function of $T_{1}$ for the Rankine cycle with the low temperature $T_{2}$=0.005 and $T_{E}=0.08$. The charge $q$=0.6,0.8,1,1.2,1.4 from top to bottom.}\label{pRankinea}
\end{figure}

\begin{figure}
\center{\subfigure[]{\label{pREffQ}
\includegraphics[width=7cm]{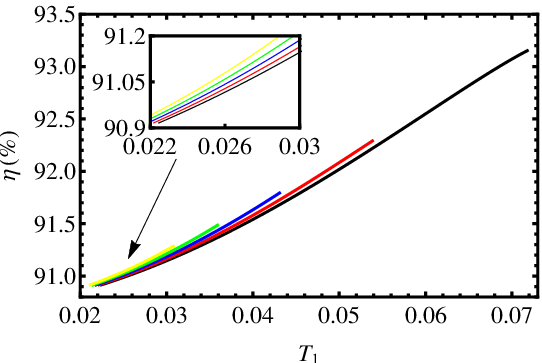}}
\subfigure[]{\label{pREffTE}
\includegraphics[width=7cm]{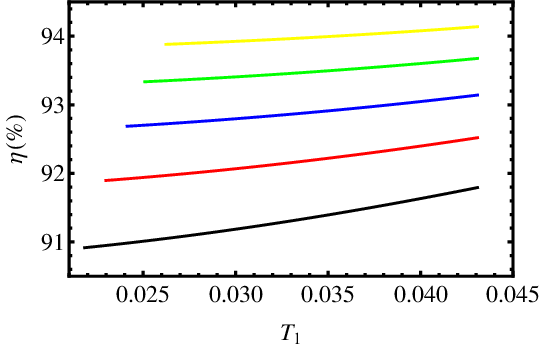}}}
\caption{Efficiency for the Rankine cycle. (a) $T_{2}$=0.005, $T_{E}=0.08$. $q$=0.6,0.8,1,1.2,1.4 from bottom to top. (b) $q$=1, $T_{2}$=0.005, and $T_{E}=$0.08, 0.09 0.1, 0.11, and 0.12 from bottom to top.}\label{pRankineb}
\end{figure}

Now, we would like to deal with the Rankine cycle, along which the steam power plant operates. We plot the heat and work in Fig.~\ref{pRankinea} with the low temperature $T_{2}$=0.005 and the temperature of point E $T_{E}=0.08$. Since point F should fall into the coexistence region, $T_{1}$ must have a minimum value larger than $T_{2}$. Both the work and heat decrease from its maximum at the minimum of $T_{1}$, and approach to a small value at the critical temperature. For a fixed $T_{1}$, the heat and work decrease with the charge $q$. The corresponding efficiency is shown in Fig.~\ref{pREffQ}. A black hole with small charge has a high critical temperature, which gives a chance for the engine to approach a high efficiency. We can also see that the efficiency increases with $q$ if $T_{1}$ is fixed. On the other hand, it is implied by the Carnot engine that, if the low temperature is fixed, the efficiency will be increased by rising its high temperature. Thus a heat engine may approach a high efficiency with increasing $T_{E}$. For this scheme, we show the efficiency in Fig.~\ref{pREffTE} with $T_{E}$=0.08, 0.09, 0.1, 0.11, and 0.12 from bottom to top. It is very clear that the efficiency increases with $T_{E}$. For example, with $T_{1}$=0.035, the efficiency $\eta$= 91.40\%, 92.22\%, 92.91\%, 93.50\%, and 93.99\% for $T_{E}$=0.08, 0.09, 0.1, 0.11, and 0.12, respectively. A much more efficiency will be achieved if $T_{E}$ is high enough.

From Figs.~\ref{pRankineWork} and \ref{pREffQ}, it is clear that, for low $T_{1}$, the amount of the work is large, while there is a poor efficiency. And for high $T_{1}$, there is a high efficiency while with a small amount of the work. Obviously, high efficiency and large amount of the work can not be approached at the same case. A comprehensive is that we can operate the engine in a not high temperature of $T_{1}$, and then to find a way to raise the efficiency. As is shown in Fig.~\ref{pREffTE}, in order to obtain a high efficiency, one can raise $T_{E}$. For a steam power plant, this mechanism is implemented with a back pressure turbine. The effect is to make the point F to fall on the coexistence curve.

\begin{figure}
\center{\subfigure[Heat for the BPR cycle]{\label{pRankineBeiHeat}
\includegraphics[width=7cm]{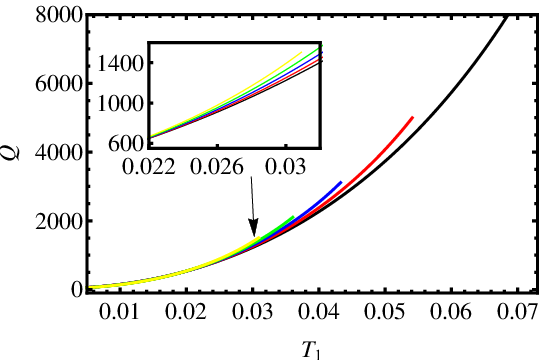}}
\subfigure[Work for the BPR cycle]{\label{pRankineBeiWork}
\includegraphics[width=7cm]{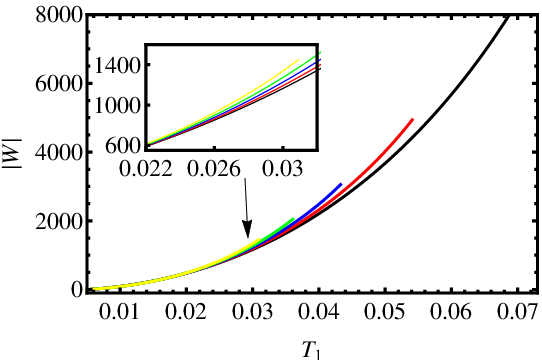}}}
\caption{(a) Heat and (b) work as functions of $T_{1}$ for the BPR cycle with the low temperature $T_{2}$=0.005. The charge $q$=0.6, 0.8, 1, 1.2, and 1.4 from bottom to top.}\label{pRankinec}
\end{figure}

\begin{figure}
\center{\subfigure[Efficiency for the BPR cycle.]{\label{pRBEffQ}
\includegraphics[width=7cm]{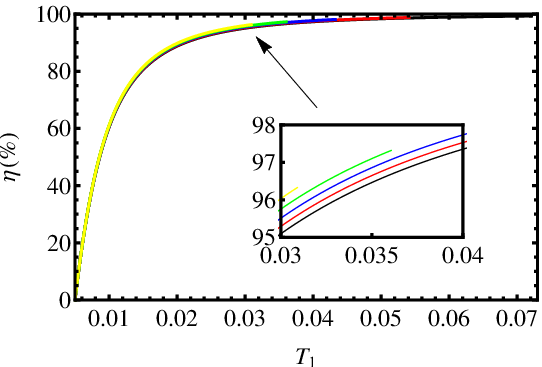}}
\subfigure[Ratio $T_{E}/T_{c}$ for the BPR cycle.]{\label{pRBTE}
\includegraphics[width=7cm]{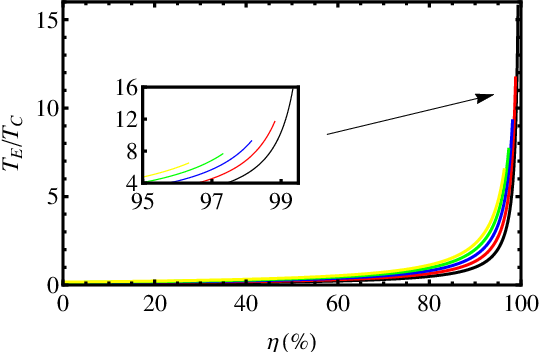}}}
\caption{Efficiency and ratio $T_{E}/T_{c}$ for the BPR cycle. Charge $q$=0.6, 0.8, 1, 1.2, and 1.4 from bottom to top. (a) Efficiency vs. $T_{1}$ with $T_{2}$=0.005 and $T_{E}=0.08$. (b) Ratio $T_{E}/T_{c}$ as a function of the efficiency, with $T_{2}$=0.005, and $T_{c}$ corresponds to the critical temperature.}\label{pRankined}
\end{figure}

Next, it is worth to examine such a cycle when a back pressure mechanism is included in. We call it the back pressure Rankine (BPR) cycle. The work and heat can be numerically obtained, and the result is displayed in Fig.~\ref{pRankinec}. It is interesting that the heat and work increase with $T_{1}$, which behaves very different from the cycle without the pressure mechanism. The efficiency also increases with $T_{1}$ shown in Fig.~\ref{pRBEffQ}. Moreover, one can also see that with the same $T_{1}$, the work and efficiency increase with the charge $q$. For this cycle, we find that the maximum work and high efficiency can be both approached at high $T_{1}$. Thus this BPR heat engine can be treated as an efficient power plant to produce the useful mechanical work. However, one should keep in mind that $T_{1}$ is not the highest temperature for the engine, while $T_{E}$ is. We show the ratio $T_{E}/T_{C}$ as a function of the efficiency $\eta$ in Fig.~\ref{pRBTE}. For small $\eta$, the ratio $T_{E}/T_{c}$ is low. However, in order to given a $\eta>$90\%, the ratio rapidly increases. Nevertheless, comparing with the standard Rankine heat engine, the BPR engine can still be an efficient power plant.

\subsection{$d\geq$5-dimensional black holes}

\begin{figure}
\includegraphics[width=8cm]{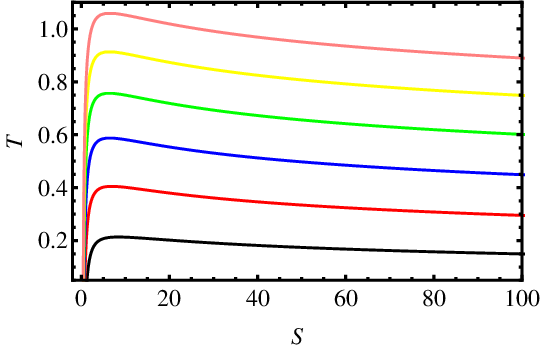}
\caption{Phase diagram in the $T$-$S$ chart with the charge $q$=1. The dimension $d$=5, 6, 7, 8, 9, and 10 from bottom to top.} \label{phts}
\end{figure}

The line element for a spherical charged AdS black hole in $d\geq$5 spacetime dimensions is
\begin{eqnarray}
 ds^{2}&=&-f(r)dt^{2}+\frac{dr^{2}}{f(r)}+r^{2}d\Omega^{2}_{d-2},\\
 f(r)&=&1-\frac{m}{r^{d-3}}+\frac{\tilde{q}^{2}}{r^{2(d-3)}}+\frac{r^{2}}{l^{2}}.
\end{eqnarray}
The parameters $m$ and $\tilde{q}$ correspond to the enthalpy and charge of the black hole,
\begin{eqnarray}
 H=\frac{d-2}{16\pi}\omega_{d-2}m,\quad
 q=\frac{\sqrt{2(d-2)(d-3)}}{8\pi} \omega_{d-2}\tilde{q},\label{d5charge}
\end{eqnarray}
where $\omega_{d-2}=2\pi^{(d+1)/2}/\Gamma((d+1)/2)$ is the volume of the unit $(d-2)$-sphere. The black hole temperature and entropy can be calculated as
\begin{eqnarray}
 T&=&\frac{d-3}{4\pi r_{h}}
    \left(1-\frac{q^{2}}{r_{h}^{2(d-3)}}+\frac{(d-1)r_{h}^{2}}{(d-3)l^{2}}\right),
        \label{d5Te}\\
 S&=&\frac{A_{d-2}}{4}=\frac{\omega_{d-2}r_{h}^{d-2}}{4}.\label{d5entropy}
\end{eqnarray}
The state equation for the black hole can be obtained with substituting Eqs. (\ref{d5charge}) and (\ref{d5entropy}) into Eq.~(\ref{d5Te})
\begin{eqnarray}
 T&=&\frac{d-3}{4\pi}\left(\frac{\omega_{d-2}}{4S}\right)^{\frac{1}{d-2}}
   \bigg[1+\frac{16\pi P}{(d-2)(d-3)}\left(\frac{4S}{\omega_{d-2}}\right)^{\frac{2}{d-2}}\nonumber\\
     &&-\frac{32\pi^{2}q^{2}}{(d-2)(d-3)\omega^{2}}\left(\frac{\omega_{d-2}}{4S}\right)^{\frac{2(d-3)}{d-2}}\bigg].\label{d5tasta}
\end{eqnarray}
The black hole enthalpy can also be expressed in terms of the charge, entropy, and pressure, as
\begin{eqnarray}
 H&=&\frac{(d-2)\omega_{d-2}}{16\pi}
   \bigg[\frac{32\pi^{2}q^{2}}{(d-2)(d-3)\omega_{d-2}^{2}}\left(\frac{4S}{\omega_{d-2}}\right)^{\frac{3-d}{d-2}}\nonumber\\
   &+&\left(\frac{4S}{\omega_{d-2}}\right)^{\frac{d-3}{d-2}}
      \left(1+\left(\frac{4S}{\omega_{d-2}}\right)^{\frac{2}{d-2}}
      \frac{16\pi P}{(d-1)(d-2)}\right)\bigg].\nonumber\\
\end{eqnarray}
This $d\geq$5-dimensional black hole system described by the state equation (\ref{d5tasta}) displays a small-large black hole phase transition, with the critical point is given by \cite{Gunasekaran},
\begin{eqnarray}
 T_{c}&=&\frac{4(d-3)^{2}}{(d-2)(2d-5)\pi v_{c}},\\
 P_{c}&=&\frac{(d-3)^{2}}{(d-2)^{2}\pi v_{c}^{2}},\\
 S_{c}&=&\frac{\omega_{d-2}}{4}(\frac{d-2}{4})^{d-2}v_{c}^{d-2},\quad\\
 v_{c}&=&\frac{4}{d-2}\left(q^{2}(d-2)(2d-5)\right)^{1/(2d-6)}, \label{critical}
\end{eqnarray}
The phase structure in the $T$-$S$ chart is plotted in Fig.~\ref{phts} for $d$=5-10 from bottom to top. Increasing the dimension $d$, the critical point of the temperature raises. Thus, the coexistence region of the small and large black holes gets larger. Based on such phase structure, we can build the maximal Carnot cycle and the Rankine cycle for the higher-dimensional charged AdS black hole.

After a simple calculation, we find that the heat, work, and the efficiency of the maximum Carnot cycle share the similar behavior as the case of $d=4$, so we will not show them.

\begin{figure}
\center{\subfigure[Work for the Rankine cycle.]{\label{phRankinea}
\includegraphics[width=7cm]{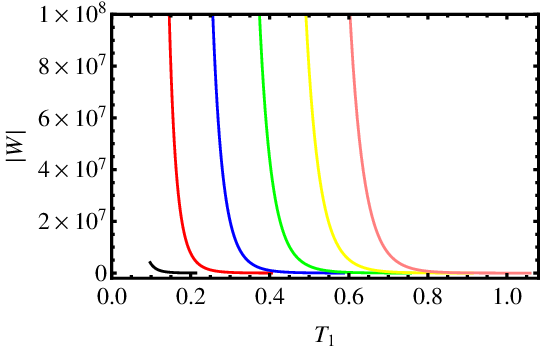}}
\subfigure[Efficiency for the Rankine cycle.]{\label{phRankineb}
\includegraphics[width=7cm]{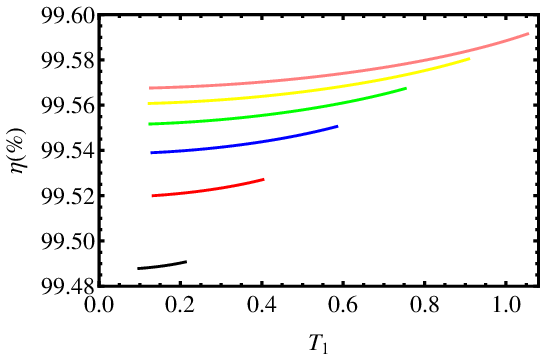}}}
\caption{Work and efficiency for the Rankine cycle with $T_{2}$=0.005, $T_{E}$=1.3, and charge $q$=1. (a) Work for the cycle. From left to right, $d$=5, 6, 7, 8, 9, and 10. (b) Efficiency for the cycle. From bottom to top, the dimension $d$=5, 6, 7, 8, 9, and 10.}\label{phRankine}
\end{figure}

\begin{figure}
\subfigure[$d$=5]{\label{phbRankinea}
\includegraphics[width=4cm]{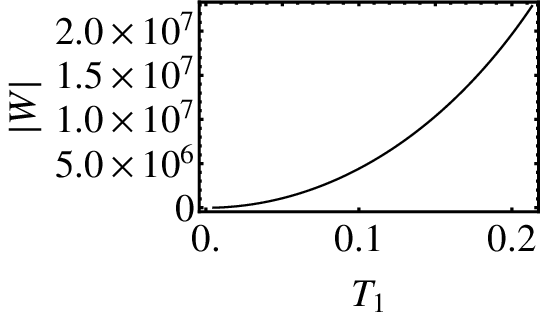}}
\subfigure[$d$=6]{\label{phbRankineb}
\includegraphics[width=4cm]{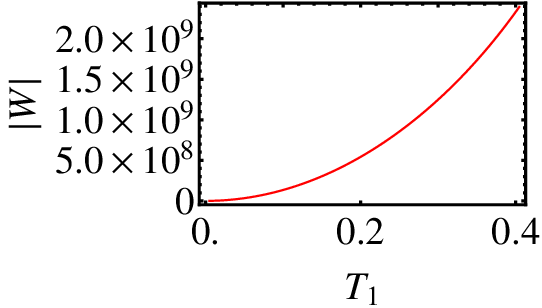}}
\subfigure[$d$=7]{\label{phbRankinec}
\includegraphics[width=4cm]{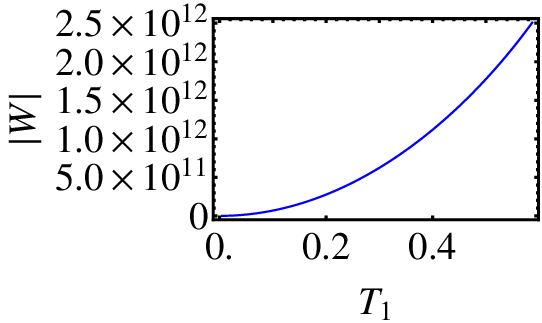}}
\subfigure[$d$=8]{\label{phbRankined}
\includegraphics[width=4cm]{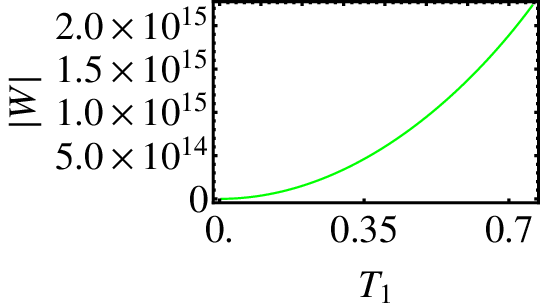}}
\subfigure[$d$=9]{\label{phbRankinee}
\includegraphics[width=4cm]{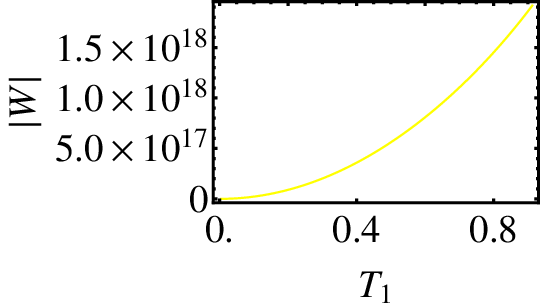}}
\subfigure[$d$=10]{\label{phbRankinef}
\includegraphics[width=4cm]{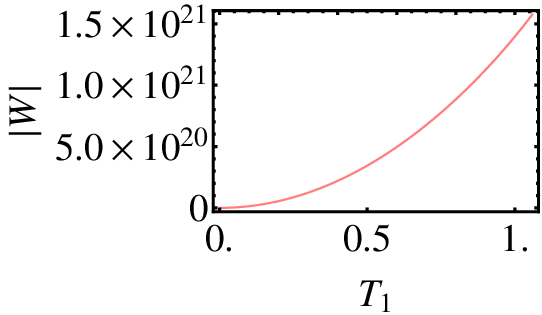}}
\caption{Work for the BPR cycle with $T_{2}$=0.005 and $q$=1. (a) $d$=5. (b) $d$=6. (c) $d$=7. (d) $d$=8. (e) $d$=9. (f) $d$=10.}\label{phbRankine}
\end{figure}

\begin{figure}
\includegraphics[width=8cm]{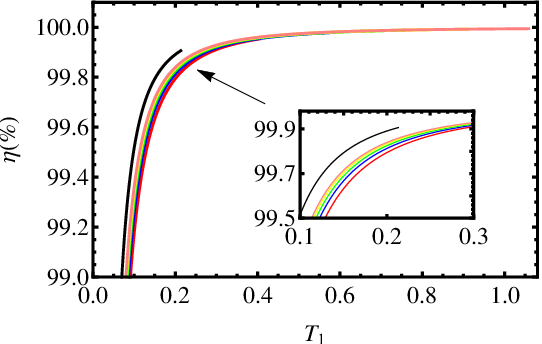}
\caption{The efficiency for the BPR cycle with $T_{2}$=0.005 and $q$=1. From top to bottom, $d$=5, 6, 7, 8, 9, and 10.} \label{pheff}
\end{figure}

For the Rankine cycle, we show its work and efficiency in Fig.~\ref{phRankine} for different dimensions. It can be seen in Fig.~\ref{phRankinea} that the work monotonically decreases with $T_{1}$ for each $d$, and it approaches to its minimum when the critical temperature is reached. The efficiency shows a small increase with $T_{1}$. For fixed $T_{1}$, $\eta$ increases with the dimension $d$. For the BPR cycle, the work is plotted against $T_{1}$ in Fig.~\ref{phbRankine} with $d$=5-10. For different $d$, the work presents a monotonically increasing behavior. And it is greatly increased with the dimension. The corresponding efficiency is shown in Fig.~\ref{pheff}. For fixed $T_{1}$, $\eta$ slightly decreases with $d$. However, it will larger than 99\% for any $d$ when $T_{1}$ approaches to the critical temperature.

One word to summarize this section is that a heat engine by a higher-dimensional black hole can be more efficient than the lower-dimensional one, especially for the BPR cycle.

\section{Reduced black hole heat engine}
\label{Reduced}

Now, It is well known that the critical phenomena of the charged AdS black hole is charge-independent in the reduced parameter space \cite{Wei4}. So here it is worth generalizing the heat engine to the reduced parameter space. At first, we define the reduced work, heat, and enthalpy as,
\begin{eqnarray}
 \tilde{Q}&=&Q/T_{c}S_{c}=\int \tilde{T}d\tilde{S},\\
 \tilde{W}&=&W/T_{c}S_{c},\\
 \tilde{H}&=&H/T_{c}S_{c}.
\end{eqnarray}
A brief calculation indicates that they are charge-independent. Then Eq.~(\ref{HW}) can be expressed as
\begin{eqnarray}
 \tilde{Q}_{1}+\tilde{Q}_{2}+\tilde{W}=0.
\end{eqnarray}
Therefore, the efficiency can be obtained through measuring the area in the reduced $\tilde{T}$-$\tilde{S}$ chart. And the reduced work $\tilde{W}$ is just the negative of the area enclosed in the cycle shown in the reduced $\tilde{T}$-$\tilde{S}$ chart. For the Rankine and BPR cycles, the efficiencies will be
\begin{eqnarray}
 \eta&=&1-\frac{\tilde{T}_{A}(\tilde{S}_{E}-\tilde{S}_{B})}
    {\tilde{H}_{\tilde{P}_{B}}(\tilde{S}_{E})-\tilde{H}_{\tilde{P}_{B}}(\tilde{S}_{B})},
\end{eqnarray}
and
\begin{eqnarray}
 \eta&=&1-\frac{\tilde{H}_{\tilde{P}_{A}}(\tilde{S}_{E})
      -\tilde{H}_{\tilde{P}_{A}}(\tilde{S}_{B})}
      {\tilde{H}_{\tilde{P}_{B}}(\tilde{S}_{E})-\tilde{H}_{\tilde{P}_{B}}(\tilde{S}_{B})},
\end{eqnarray}
respectively.

\subsection{$d=4$-dimensional black holes}

\begin{figure}
\center{\subfigure[Reduced $\tilde{T}$-$\tilde{S}$ phase diagram.]{\label{preduceda}
\includegraphics[width=7cm]{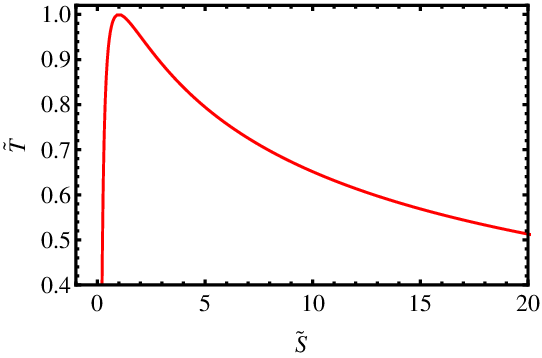}}
\subfigure[Reduced work for the Carnot heat engine]{\label{preducedb}
\includegraphics[width=7cm]{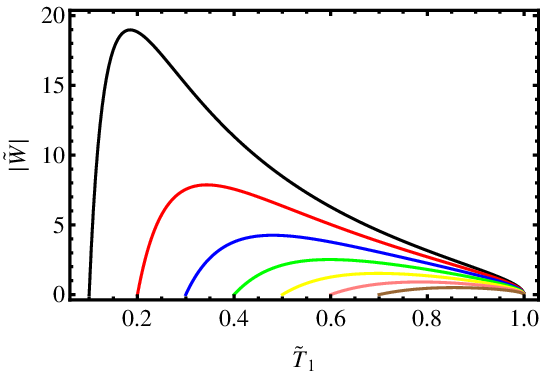}}}
\caption{(a) Reduced $\tilde{T}$-$\tilde{S}$ phase diagram for $d$=4. (b) Reduced work for the Carnot heat engine by the four-dimensional charged black hole with the low temperature $\tilde{T}_{2}$=0.1, 0.2, 0.3, 0.4, 0.5, 0.6, and 0.7, respectively, from top to bottom.}\label{preduced}
\end{figure}

For the $d=4$-dimensional black hole system, the reduced state equation can be expressed as
\begin{eqnarray}
 \tilde{T}=\frac{3}{8}\left(\tilde{P}\sqrt{\tilde{S}}+\frac{2}{\sqrt{\tilde{S}}}
     -\frac{1}{3\tilde{S}^{3/2}}\right).
\end{eqnarray}
It is clear that such state equation is charge-independent. Constructing the equal are law on each isobaric line, the phase transition point will be determined in the reduced $\tilde{T}$-$\tilde{S}$ chart, which is obviously charge independent. In Ref.~\cite{Lan}, the author obtained an analytical formula of the coexistence curve in the $P$-$T$ chart for the four-dimensional black hole by constructing the equal-area law in the reduced parameter space. The result shows that the formula is independent of the black hole charge. Similarly, the coexistence curve in the reduced $\tilde{T}$-$\tilde{S}$ chart can also be obtained with the method, or expressed Eq.~(\ref{tscoe}) in the reduced parameter space as
\begin{equation}
 \tilde{T}^{2}=1+\cos\left(3\arccos\Big(\frac{2+\tilde{S}-\sqrt{3+6\tilde{S}}}{2\tilde{S}}\Big)\right).
\end{equation}
The phase structure is shown in Fig.~\ref{preduceda}. Similar to that in the ordinary parameter space, there are three regions. It is worthwhile to note that the critical point is casted into the (1, 1) point. Based on such phase structure, the maximum Carnot cycle and the Rankine cycle can be built.

First, let us consider the maximum Carnot cycle. Set the low temperature $\tilde{T}_{2}$=0.1, 0.2, 0.3, 0.4, 0.5, 0.6, and 0.7, we show the work in Fig.~\ref{preducedb}. One can see that the behavior of the work is similar for different values of $\tilde{T}_{2}$, and the maximum of the work decreases with $\tilde{T}_{2}$. So, decreasing $\tilde{T}_{2}$ will give a maximum work. On the other hand, the efficiency for the maximum Carnot cycle is $\eta=1-\frac{T_{2}}{T_{1}}=1-\frac{\tilde{T}_{2}}{\tilde{T}_{1}}$.

\begin{figure}
\center{\subfigure[Work]{\label{pReducedRankineEffa}
\includegraphics[width=7cm]{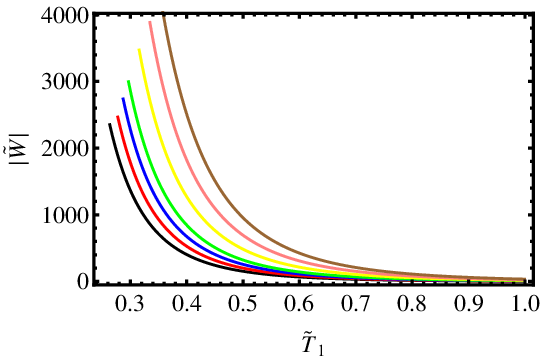}}
\subfigure[Efficiency]{\label{pReducedRankineEffb}
\includegraphics[width=7cm]{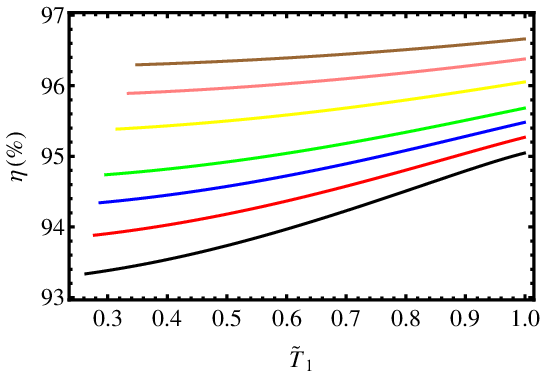}}}
\caption{Work and efficiency for the reduced Rankine cycle with low temperature $\tilde{T}_{2}$=0.05, and $\tilde{T}_{E}$=1.1, 1.2, 1.3, 1.4, 1.6, 1.8, and 2.0 from bottom to top.}\label{pReducedRankineEff}
\end{figure}

\begin{figure}
\center{\subfigure[Work]{\label{pReducedRankineEffabpr}
\includegraphics[width=7cm]{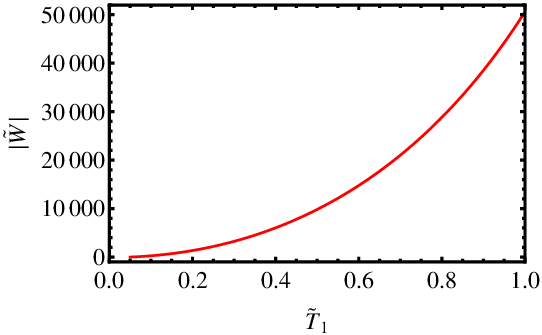}}
\subfigure[Efficiency]{\label{pReducedRankineEffbbpr}
\includegraphics[width=7cm]{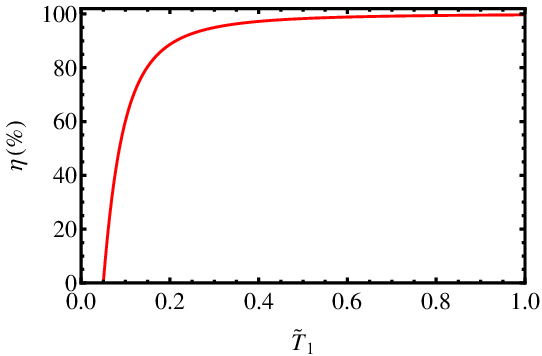}}}
\caption{Work and efficiency for the reduced BPR cycle with low temperature $\tilde{T}_{2}$=0.05. (a) Work. (b) Efficiency.}\label{pReducedRankineEffbpr}
\end{figure}

For the reduced Rankine cycle, the reduced work and efficiency are given in Fig.~\ref{pReducedRankineEff}. It is clear that the reduced work decreases, while the efficiency increases with $\tilde{T}_{1}$. For fixed $\tilde{T}_{1}$, the work and efficiency both increase with $\tilde{T}_{E}$. Similar to that in the ordinary parameter space, the work approaches to a minimum, and the efficiency approaches to a maximum when the $\tilde{T}_{1}$ has a value of the critical temperature. In order to achieve the large amount of the work and the high efficiency at the same time, one way is to introduce the back pressure mechanism. We show the reduced work and the efficiency in Fig.~\ref{pReducedRankineEffbpr} for the reduced BPR cycle. And we can see that the maximum work and the high efficiency can be both achieved at high $\tilde{T}_{1}$.

\begin{figure}
\includegraphics[width=8cm]{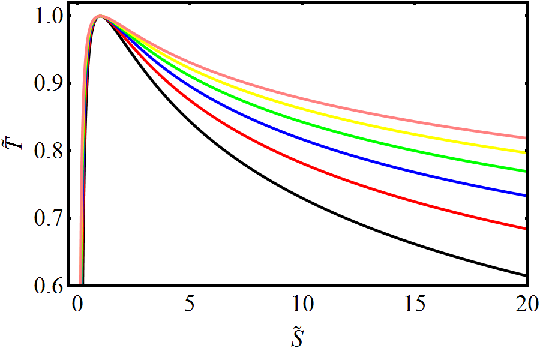}
\caption{Reduced $\tilde{T}$-$\tilde{S}$ phase diagram for $d$=5-10 from bottom to top.} \label{tsss}
\end{figure}

\begin{figure}
\center{\subfigure[Work]{\label{pReducedRankineEff2a}
\includegraphics[width=7cm]{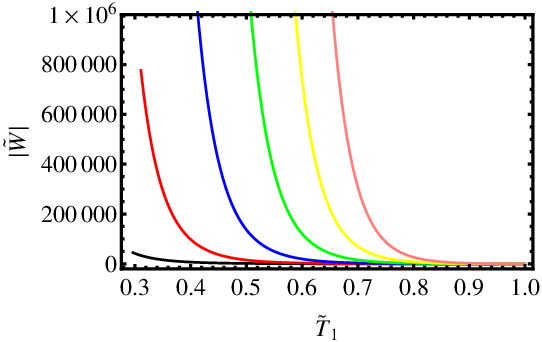}}
\subfigure[Efficiency]{\label{pReducedRankineEff2b}
\includegraphics[width=7cm]{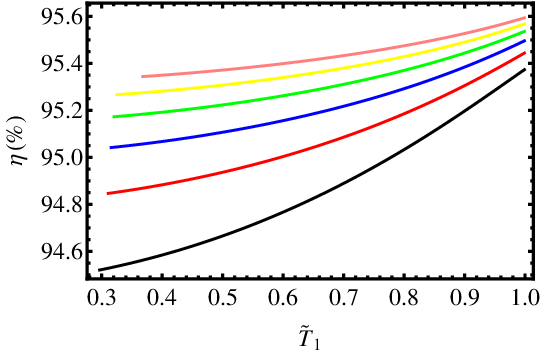}}}
\caption{Work and efficiency for the reduced Rankine cycle with low temperature $\tilde{T}_{2}$=0.05 and $\tilde{T}_{E}=1.2$ for $d$=5-10 from bottom to top. (a) Work. (b) Efficiency.}\label{pReducedRankineEff2}
\end{figure}

\begin{figure}
\includegraphics[width=8cm]{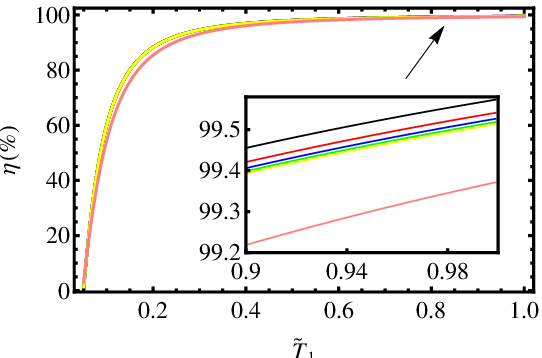}
\caption{Efficiency for the reduced BPR cycle with $\tilde{T}_{2}$=0.05 for $d$=5-10 from top to bottom.} \label{pheff2}
\end{figure}

\subsection{$d\geq$5-dimensional black holes}

There also exists the small-large black hole phase transition in the higher spacetime dimensions. In the reduced parameter space, the state equation reads
\begin{eqnarray}
 \tilde{T}=\frac{\tilde{S}^{\frac{5-2d}{d-2}}}{4(d-3)(d-2)}
   \bigg(-1&+&(2d-5)\tilde{S}^{2}\bigg((d-3)\tilde{P}\nonumber\\
      &+&(d-2)\tilde{S}^{\frac{2}{2-d}}\bigg)\bigg),
\end{eqnarray}
which is also independent of the black hole charge. Constructing the equal-area law, one can determine the phase transition point. And let the reduced pressure vary from 0 to 1, the coexistence curve will be obtained. Different from the $d$=4 case, there is no analytical formula of the coexistence curve for the higher-dimensional black hole. However, the highly accurate fitting formula for $d$=5-10 dimensional black hole has been presented in Ref.~\cite{Wei4}. Using them, we plot the coexistence curve in the reduced $\tilde{T}$-$\tilde{S}$ chart in Fig.~\ref{tsss}, from which one can see that the coexistence curve of small black hole is almost coincidence with each other for any dimension, while the coexistence curve of large black hole increases with $d$, which enlarges the coexistence region.

For the reduced Rankine cycle with low temperature $\tilde{T}_{2}$=0.05 and $\tilde{T}_{E}=1.2$, we plot the reduced work and efficiency in Fig.~\ref{pReducedRankineEff2} for $d$=5-10. The work decreases, while the efficiency increases with $\tilde{T}_{1}$. Another significant result is that the work and efficiency both increase with the number of spacetime dimensions $d$ for a fixed $\tilde{T}_{1}$. And the work shows a significant growth with $d$.

For the reduced BPR cycle with low temperature $\tilde{T}_{2}$=0.05, we list the reduced work in Table \ref{parameters}. For the same $d$, the reduced work increases with $\tilde{T}_{1}$. And for fixed $\tilde{T}_{1}$, it increases rapidly with $d$. The efficiency is also plotted in Fig.~\ref{pheff2}. For high $T_{1}$, the efficiency is larger than 99\% for each $d>5$. Thus the heat engine by a higher-dimensional black hole will be a power plant of a large amount of work and high efficiency.

\begin{widetext}
\begin{center}
\begin{table}[h]
\begin{center}
\begin{tabular}{c c c c c c c c}
  \hline
  \hline
   &$\tilde{T}_{1}$=0.6 &$\tilde{T}_{1}$=0.7 & $\tilde{T}_{1}$=0.8 & $\tilde{T}_{1}$=0.9 & $\tilde{T}_{1}$=0.95 & $\tilde{T}_{1}$=0.98 \\
   $d$=5 & 1.98$\times 10^{5}$ & 2.75$\times 10^{5}$ & 3.68$\times 10^{5}$
     & 4.80$\times 10^{5}$ & 5.44$\times 10^{5}$ & 5.86$\times 10^{5}$ \\
   $d$=6 & 3.42$\times 10^{6}$ & 4.71$\times 10^{6}$ & 6.24$\times 10^{6}$
     & 8.05$\times 10^{6}$ & 9.08$\times 10^{6}$ & 9.73$\times 10^{6}$ \\
   $d$=7 & 6.37$\times 10^{7}$ & 8.73$\times 10^{7}$ & 1.15$\times 10^{8}$
     & 1.48$\times 10^{8}$ & 1.66$\times 10^{8}$ & 1.78$\times 10^{8}$ \\
   $d$=8 & 1.22$\times 10^{9}$ & 1.67$\times 10^{9}$ & 2.20$\times 10^{9}$
     & 2.81$\times 10^{9}$ & 3.15$\times 10^{9}$ & 3.37$\times 10^{9}$ \\
   $d$=9 & 2.35$\times 10^{10}$ & 3.21$\times 10^{10}$ & 4.22$\times 10^{10}$
     & 5.38$\times 10^{10}$ & 6.03$\times 10^{10}$ & 6.44$\times 10^{10}$ \\
  $d$=10 & 4.83$\times 10^{10}$ & 6.60$\times 10^{10}$ & 8.65$\times 10^{10}$
     & 1.10$\times 10^{11}$ & 1.23$\times 10^{11}$ & 1.32$\times 10^{11}$ \\
  \hline\hline
\end{tabular}
\end{center}
\caption{Reduced work for the BPR cycle with $\tilde{T}_{2}$=0.05.}\label{parameters}
\end{table}
\end{center}
\end{widetext}

\section{Conclusions and discussions}
\label{Conclusion}

In this paper, we have studied the heat engine operating along the maximal Carnot cycle and Rankine cycle by a charged AdS black hole. The working substance is identified with the black hole molecules. A remarkable feature of the study is that the small-large black hole phase transition is included in. For example, we require the working substance always in the coexistence phase for the maximal Carnot cycle. For the Rankine cycle, the substance will encounter a phase transition during each cycle.

At first, we briefly reviewed the first law for the black hole system. The two thermodynamic cycles were introduced. By means of the first law and the equal-area law, we showed a compact form of the efficiency for the heat engine.

Then we considered the specific black hole heat engines. For the $d$=4-dimensional case, we showed the phase structure in Fig.~\ref{pts}. It was found that the peak of the coexistence decreases with the charge $q$ leading to the shrink of the coexistence phase. Based on the phase structure, the maximal Carnot cycle and Rankine cycle were examined in detailed. Our general process is fixing the low temperature and leaving the high temperature freely varies from the low temperature to the critical temperature. For the maximal Carnot cycle, the result shows that the work firstly increases, and then decreases to zero at the critical temperature. While the heat displays a monotonically decreasing behavior. For the Rankine cycle, both the heat and work decrease with $T_{1}$. And the efficiency increases with $T_{1}$, while decreases with the charge $q$. A significant feature for the Rankine cycle is that the maximal values of the work and efficiency can not be achieved at the same case. One alternative is increasing the temperature $T_{E}$, which can be implemented with introducing the back pressure mechanism. For such BPR cycle, the maximum work and efficiency can be obtained at high $T_{1}$. The price is increasing the highest temperature $T_{E}$ with which the engine operating.

For the $d\geq$5-dimensional black hole, the critical temperature raises with the spacetime dimensions $d$ with fixed $q$. All the results of the Rankine cycle and BPR cycle show that the work rapidly increases with $d$ and the efficiency increases to approach its bounded value. Thus one gets the result that the higher-dimensional black hole heat engine can produce much more useful mechanical work, and can act as a more efficient power plant.

Furthermore, since the critical phenomenon of the charged AdS black holes in the reduced parameter space is charge-independent, we explored the black hole engine in such parameter space. The compact formulas of the efficiency were also given for the Rankine cycle with or without the back pressure mechanism. The study confirms the similar result in the ordinary parameter space. In particular, as shown in Fig.~\ref{pReducedRankineEff2a} and Table~\ref{parameters}, the reduced work rapidly increases with the number of the spacetime dimensions $d$.

A direct consequence of this article is that high dimensional black hole heat engine can perform as an efficient engine and produce a large amount of mechanical work than the low dimensional one. Such mechanical work may be a possible source to continuously accelerate the particle near the black hole. If this is true, black hole power jet, as well as other high energy phenomena, may be much more violent in high spacetime dimension. On the other hand, according to the AdS/CFT dual, it is also interesting to explore the holographic dual description of such thermodynamic cycle in the large $N$ field theory.

\section*{Acknowledgements}
This work was supported by the National Natural Science Foundation of China (Grants No. 11675064, No. 11522541,  No. 11375075, and No. 11205074), and the Fundamental Research Funds for the Central Universities (No. lzujbky-2016-121 and lzujbky-2016-k04).

\end{document}